\documentclass{PoS}

\title{Path Integral Monte-Carlo method for relativistic quantum systems}

\ShortTitle{Path Integral Monte-Carlo method for relativistic
quantum systems}

\author{\speaker{Oleg Pavlovsky}\thanks{This work is supported in part by RFBR Grant No.
14-02-01261. Numerical calculations were performed at
Supercomputing Center of the Moscow State University [9].}\\
        Institute for Theoretical Problems of Microphysics, Moscow State University and Institute for Theoretical and Experimental Physics, Moscow, Russia\\
        E-mail: \email{ovp@goa.bog.msu.ru}}

\author{Alexander Ivanov\\
        Department of Physics, Moscow State University, Moscow, Russia\\
        E-mail: \email{ivanov.as@physics.msu.ru}}

\author{Alexander Novoselov\\
         Institute for Theoretical Problems of Microphysics, Moscow State University and Institute for Theoretical and Experimental Physics, Moscow, Russia\\
         E-mail: \email{novoselov@goa.bog.msu.ru}}

\abstract{  Relativistic generalization of Path Integral
Monte-Carlo method has been proposed and  some possible
applications have been discussed. }

\FullConference{The 32nd International Symposium on Lattice Field Theory\\
                 23-28 June, 2014\\
                 Columbia University New York, NY}

\begin{document}

\section*{Introduction}

This work is devoted to the generalization of Path Integral Monte
Carlo approach \cite{Ceperley} in the case of the relativistic
systems. There are many physical problems connected with
simulations of relativistic quantum mechanical systems.
Relativistic corrections play very essential role in  physics of
the atomic systems with heavy elements due to the strong
interaction potentials. One may find the problems with simulations
of relativistic quantum systems in the nuclear physics, physics of
hadron structure, quark-gluon plasma and in relativistic
astrophysics. Recently an another interesting application of the
relativistic quantum mechanics has arisen. There are so called
(Pseudo)Relativistic Condensed Matter systems and one of the must
interesting example of such systems is Graphene
(\cite{Novoselov:04:1}).

The correct formulation of the many-body relativistic
quantum-mechanical  problem has some well-known difficulties. The
kinetic and potential part of the Hamiltonian must be invariant
under Lorentz transformations. The kinetic part of the Hamiltonian
can be formulated in Lorentz-invariant form relatively easy but
relativistic formulation of interaction requires the quantum field
theory approach in general case. In this work we have considered
the problem with the instantaneous interaction in the Hamiltonian.
This approximation works very well in Relativistic Quantum
Chemistry and in Physics of the (Pseudo)Relativistic Condensed
Matter systems like Graphene.

In our work we have discussed the basic concepts of the
relativistic generalization of the Path Integral approach. The
main observation is that in relativistic case the statistical
weight of the ``Path'' is not an exponent of some classical action
as it is in the case of the non-relativistic systems due to the
well-known Feynman-Kac formula. In spite of this fact, the
relativistic statistical weight of the ``Path'' is well defined
positive function and can be easily used in numerical
calculations. In this work we test our approach on the simple
academic problem - Relativistic Oscillator problem. This system
gives us good opportunity for testing of the Relativistic Path
Integral Monte-Carlo approach because one can study this system by
using of standard Schrodinger equation approach in the momentum
space. We will compare the results of these two approaches.

\section{Non-relativistic and ultra-relativistic limits for relativistic quantum particles}

Let us consider the quantum system with following Hamiltonian
$$
H = T(p) + V(q) = \sqrt{p ^ 2 + m ^ 2} + V(q),
$$
where $q$ and $p$ is the coordinate and the momentum of the
particle. $T(p)$ is the kinetic energy with rest mass and $V(q)$
is the potential energy.

In non-relativistic limit $ m ^ 2 \gg p ^ 2$ one gets
$$
T(p) = \sqrt{p ^ 2 + m ^ 2} = m (1 + \frac{p ^ 2}{2 m ^ 2} + O
\Bigl ( \Bigl ( \frac{p}{m} \Bigr ) ^ 4 \Bigr ) ) \approx m +
\frac{p ^ 2}{2 m} ,
$$
Typical values of momentum $p$ are dependent on the potential
$V(q)$. If the interaction is strong, the values of momentum $p$
can be very large. In this case we are not able to use the
non-relativistic Hamiltonian and relativistic corrections are very
essential.

As an example, in our work we have considered the Harmonic
oscillator potential
\begin{equation}
V (q) = \frac{1}{2} m \omega ^ 2 q ^ 2 .
\end{equation}

In non-relativistic case the virial theorem gives us the following
relation between kinetic and potential energy
$$
\Bigl \langle \frac{p ^ 2}{2 m} \Bigr \rangle = \langle
\frac{1}{2} m \omega ^ 2 q ^ 2 \rangle \, \mbox{ or } \, \langle
T(p) - m \rangle = \langle V (q) \rangle.
$$
Using this expression we can formulate  non-relativistic condition
for this system in terms of $m$ and $\omega$:
$$
\langle p ^ 2 \rangle \sim m \omega \, \Rightarrow \, m \gg
\omega.
$$
It means that in non-relativistic case we consider heavy particles
and soft potentials.

Similarly let us consider ultra-relativistic limit ($ \langle p ^
2 \rangle \gg m ^ 2$). The kinetic energy in this case is follows:
$$
T (p) = \sqrt{p ^ 2 + m ^ 2} = |p| \Bigl ( 1 + \frac{m ^ 2}{2 p ^
2} + O \Bigl ( \Bigl ( \frac{m}{p} \Bigr ) ^ 4 \Bigr ) \Bigr )
\approx |p|.
$$
Let us consider the Hamiltonian in zero order approximation:
\begin{equation}\label{hamiltinianulrel}
H = |p| + \frac{1}{2} m \omega ^ 2 q ^ 2.
\end{equation}
In this case one can solve Schrodinger equation for this
Hamiltonian in momentum representation and find energy of ground
state and corresponding wave function. The virial theorem for this
Hamiltonian gives us the relation between kinetic and potential
energy
$$
\langle T (p) \rangle = 2 \langle V (q) \rangle = \frac{2 \lambda
_ 0}{3} (m \omega ^ 2) ^ {1 / 3},
$$
where $\lambda _ 0 = 0.808617\dots$. One obtains that $\langle |p|
\rangle \sim (m \omega ^ 2) ^ {1 / 3}$ and we have opposite
expression for mass and frequency for ultra-relativistic case:
$$
\omega \gg m.
$$
So, ultra-relativistic limit in this problem is reached in the
case of small mass and strong potentials.

\section{Path integral approach for the relativistic quantum particles}

 Relativistic generalization of the Path Integral approach for
 quantum mechanical systems has a long history \cite{Fiziev},
 \cite{Redmount}. Today this approach becoming more and more popular
 and find its application in high-energy physics \cite{Filinov1},
 \cite{Filinov2}.

Now let us consider relativistic generalization of the Path
Integral approach.  Matrix element of density matrix in the
position representation is
$$
    \rho (q, q '; \beta) = \langle q | e ^ {-\beta H} | q '
    \rangle
    = \int \dots \int dq _ 1 dq _ 2 \dots dq _ {N - 1} \rho(q , q _ 1; \tau) \rho (q _ 1, q _ 2; \tau) \dots \rho(q _ {N - 1}, q '; \tau).
$$
Operator of kinetic energy is diagonal in the momentum
representation and operator of potential energy is diagonal in the
position representation. We can separate kinetic and potential
energy if $\tau$ is small:
$$
    \rho (q _ 0, q _ 2; \tau) \approx \int dq _ 1 \langle q _ 0 | e ^ {-\tau T} | q _ 1 \rangle \langle q _ 1 | e ^ {-\tau V} | q _ 2 \rangle.
$$
Potential energy is diagonal in this representation and for
kinetic energy part one obtains
$$
    \langle q _ 0 | e ^ {-\tau T} | q _ 1 \rangle = \int dp dp ' \delta (p - p') \langle q _ 0 | p \rangle \langle p ' | q _ 1 \rangle e ^ {-T (p) \tau}
 = \int \frac{dp}{2 \pi} e ^ {-T (p) \tau -i p (q _ 0 - q _ 1)}.
$$
So to build the Path Integral we must take this integral over
momentum. One can calculate integral over momentum with $T(p)=
\sqrt{p ^ 2 + m ^ 2}$
$$
\langle q _ 0 | e ^ {-\tau T} | q _ 1 \rangle = \frac{m \tau}{\pi
\sqrt{\tau ^ 2 + (q _ 1 - q _ 0) ^ 2}} K _ 1 (m \sqrt{\tau ^ 2 +
(q _ 1 - q _ 0) ^ 2}) ,
$$
where $K _ 1 (*)$ is modified Bessel function of first order. The
general expression for matrix element is the following
\begin{equation}\label{matrixelem}
\rho (q '', q '; \tau) = \langle q '' | e ^ {-\tau (T(p) + V(q))}
| q ' \rangle = \frac{m}{\pi \sqrt{1 + \Bigl ( \frac{q '' - q
'}{\tau} \Bigr ) ^ 2}} K _ 1 \Bigl [ m \tau \sqrt{1 + \Bigl (
\frac{q '' - q '}{\tau} \Bigr ) ^ 2} \Bigr ] e ^ {-\tau V(q ')} .
\end{equation}

One can find the multi-dimensional generalization of the
expression (\ref{matrixelem})
\begin{equation}\label{StatIntd}
    \rho (q '', q '; \tau)  =   \Bigl ( \frac{m \tau}{\pi \sqrt{\tau ^ 2 + ({\bf{q ''}} - {\bf{q '}}) ^ 2}} \Bigr ) ^ {(d + 1) / 2} \frac{K _ {(d + 1) / 2} (m \sqrt{\tau ^ 2
    + ({\bf{q ''}} - {\bf{q '}}) ^ 2})}{(2 \tau) ^ {(d - 1) / 2}} \exp \Bigl ( -  \tau V(q ')  \Bigr ) ,
\end{equation}
where $d$ is  the space dimensions.

Unfortunately in this relativistic case we have no interpretation
of expressions (\ref{matrixelem}) and (\ref{StatIntd}) in terms of
some classical action as we have in non-relativistic case in
Feynman-Kac formula. Nevertheless, this statistical weighs of the
``Path'' are the positively definite values. It means that we can
use these expressions for the construction of the Path Integral
Monte-Carlo Algorithms.

\section{Path Integral Monte-Carlo Metropolis Algorithm}

For Path Integral Monte-Carlo Metropolis algorithm we should know
part of the density matrix which corresponds to fixed point $q _
i$. Discussion about this method and all proofs one can find in
$\cite{Creutz}$. Using ($\ref{matrixelem}$) we can write
$$
\pi (q _ i) = \frac{m ^ 2 K _ 1 \Bigl [ m \tau \sqrt{1 + \Bigl (
\frac{q _ i - q _ {i - 1}}{\tau} \Bigr ) ^ 2} \Bigr ] K _ 1 \Bigl
[ m \tau \sqrt{1 + \Bigl ( \frac{q _ {i + 1} - q _ i}{\tau} \Bigr
) ^ 2} \Bigr ]}{\pi ^ 2 \sqrt{1 + \Bigl ( \frac{q _ i - q _ {i -
1}}{\tau} \Bigr ) ^ 2} \sqrt{1 + \Bigl ( \frac{q _ {i + 1} - q _
i}{\tau} \Bigr ) ^ 2}} e ^ {-\tau V(q _ i)}.
$$
So to calculate Path Integral we construct Markov chain which has
equilibrium state for fixed $q _ i$ proportional to $\pi (q _ i)$.
In Metropolis algorithm we can split transition probability
$$
P(q _ i \to q _ i ') = T (q _ i \to q _ i ') A (q _ i \to q _ i ')
,
$$
where $T (q _ i \to q _ i ')$ is sampling distribution and $A (q _
i \to q _ i ')$ is acceptance probability
$$
A (q _ i \to q_ i ') = \mbox{min} \Bigl [1, \frac{T(q _ i ' \to q
_ i) \pi (q _ i ')}{T(q _ i \to q _ i ') \pi (q _ i)} \Bigr ]
$$
If we know density matrix in the position representation, we can
calculate expressions for average of any observables. The
expression for kinetic energy in the relativistic case is the
following (see $\cite{Rothe}$):
\begin{equation}\label{meankineticgenrel}
\langle T(p) \rangle = \Bigl \langle \frac{m \tau}{\sqrt{\tau ^ 2
+ (\Delta q) ^ 2}} \frac{K _ 0 (m \sqrt{\tau ^ 2 + (\Delta q) ^
2})}{K _ 1 (m \sqrt{\tau ^ 2 + (\Delta q) ^ 2})} + \frac{\tau ^ 2
- (\Delta q) ^ 2}{\tau (\tau ^ 2 + (\Delta q) ^ 2)} \Bigr \rangle.
\end{equation}
And total energy can be calculated as follows:
$$
\langle E(p, q) \rangle = \langle T(p) + V(q) \rangle.
$$

\section{One dimensional test: Relativistic Oscillator}

For the testing of the Relativistic PIMC approach let us consider
the simple academic problem - Relativistic Oscillator in one space
dimension. This problem can be studied  by using of the standard
Schrodinger equation approach in the momentum space and we will
compare the results of this two approaches.

On Fig. 1 one can see the results of the PIMC simulations for
kinetic and potential energy of the Relativistic Oscillator ground
state. For these simulations we have used 1500 statistically
independent configurations. In this calculation we have fixed the
value of $\omega=1$ and have changed the value of mass $m$. In
this case the ultra-relativistic limit is reached at  $m \ll 1$
while non-relativistic limit is reached at $m \gg 1$
correspondingly.

\begin{figure}
\hspace*{-0.0 cm}\centering
\includegraphics[width=0.63\textwidth]{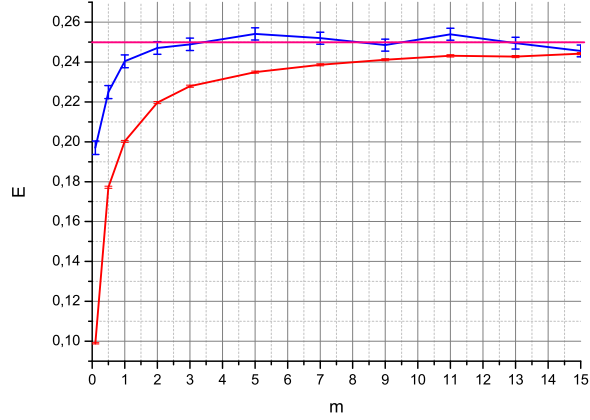}
\vspace{-0pt} \caption{PIMC simulations for kinetic (blue line)
and potential (red line) energy of the Relativistic Oscillator
ground state. $\omega=1$. } \label{fig1}
\end{figure}

The virial theorem predicts the following expressions for kinetic
and potential energy
$$
\langle T - m \rangle_{\mbox{non-rel}} = \langle V
\rangle_{\mbox{non-rel}}= 0.25
$$
at the non-relativistic limit and
$$
\langle T \rangle_{\mbox{ultra-rel}}= \langle V
\rangle_{\mbox{ultra-rel}}/2
$$
at the ultra-relativistic limit. Our PIMC results are in very good
agreement with both these expressions.

\begin{figure}
\hspace*{-0.0 cm}\centering
\includegraphics[width=0.63\textwidth]{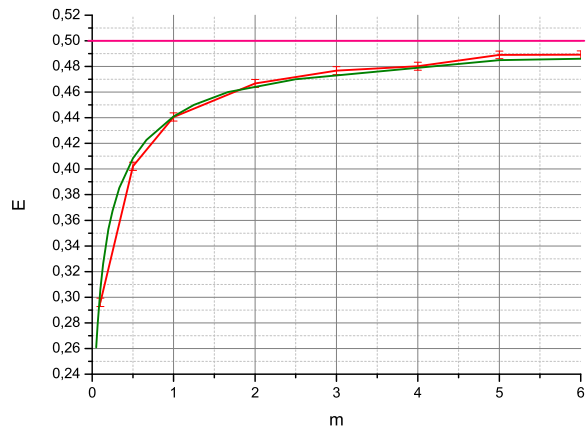} \vspace{-0pt} \caption{PIMC simulations for
total energy of the Relativistic Oscillator ground state.
$\omega=1$. The green line is the results of Schrodinger equation
approach.} \label{fig2}
\end{figure}

On Fig. 2 one can find  the results of PIMC simulations of the
total energy of the ground state at $\omega=1$. In
non-relativistic limit the energy of the ground state equals to
$\omega/2=0.5$. The green line is the result of Schrodinger
equation approach. One can see that PIMC results are in the good
agreement the results of the Schrodinger equation approach.

\begin{figure}
\hspace*{-0.0 cm}\centering
\includegraphics[width=0.63\textwidth]{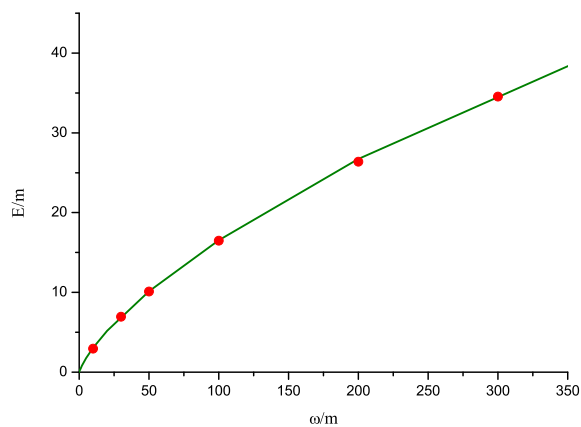}
\vspace{-0pt} \caption{Total energy of the Relativistic Oscillator
ground state for different $\omega$ and $m$. The green line is the
result of Schrodinger equation approach. Red points are the
results of PIMC approach. } \label{fig3}
\end{figure}

On Fig. 3 one can see the test of the PIMC method for strong
potentials. In this calculation one can fix the value of the mass
(for example, $m=1$) and change the value of $\omega$. The green
line is again the results of Schrodinger equation approach and red
points are the PIMC results. In this analyzing we have found out
that PIMC method can be applied for very strong potentials.

\section*{Discussion of the results and conclusion}

The main goal of our work is to construct relativistic
generalization of the PIMC method. This method we plan to use for
investigation of the properties of the relativistic quantum
systems with instantaneous interactions between the particles.
Last fact gives us possibility to  avoid the problem with
correctness of the many-body quantum-mechanical interaction. For
example, this approach could be used for studies the
thermodynamical and transport properties of graphene which are
very essential for future technological application. Of course,
there are many interesting applications one can find in
Relativistic Quantum Chemistry too.

For testing of our approach, we study simple one-dimensional
system with quadratic external potential - relativistic
oscillator. This system gives us the good possibility for testing
our approach because just relativistic oscillator can be studied
by using Schrodinger equation in momentum space. The comparison
the results of these two approaches have shown that our
relativistic generalization of PIMC method can be used for
investigation of quantum systems  which contain the relativistic
particles.


\begin{thebibliography}{25}

\bibitem{Ceperley}
    D. M. Ceperley \emph{Path integrals in the theory of condensed helium}, \emph{Rev.Mod.Phys.} \textbf{67} (1995) 279-355

\bibitem{Novoselov:04:1}
K.\,S. Novoselov, A.\,K. Geim, S.\,V. Morozov, D. Jiang, Y. Zhang,
S.\,V. Dubonos, I.\,V. Grigorieva, and A.\,A. Firsov,
\emph{Electric Field Effect in Atomically Thin Carbon Films},
\emph{Science} \textbf{306} (2004) 666.

\bibitem{Fiziev}
P. P. Fiziev, \emph{Relativistic Hameltonians wit square-roofs in
the Path Integral formaism}, \emph{Theor. Math. Phys.} {\bf 62}
(1985) 186.

\bibitem{Redmount}
I. H. Redmount and W.-M. Suen, \emph{Path integration in
relativistic quantum mechanics}, \emph{ Int. J. Mod. Phys. A} {\bf
08} (1993) 1629.

\bibitem{Filinov1}
V.S. Filinov, Yu.B. Ivanov, V.E. Fortov, M. Bonitz, and P.R.
Levashov, \emph{Color path-integral Monte-Carlo simulations of
quark-gluon plasma: Thermodynamic and transport properties},
\emph{Phys.Rev. C} {\bf 87} (2013) 035207.

\bibitem{Filinov2}
V.S. Filinov, M. Bonitz, Y.B. Ivanov, M. Ilgenfritz, and V.E.
Fortov, \emph{Thermodynamics of the quark-gluon plasma at finite
chemical potential: color path integral Monte Carlo results},
arXiv:1408.5714

\bibitem{Creutz}
    M. Creutz and B.A. Freedman,  \emph{A statistical approach to quantum mechanics}, \emph{Ann. Phys.} {\bf 132} (1981) 427.

\bibitem{Rothe}
    H. J. Rothe \emph{Lattice Gauge Theories: An Introduction} (3rd Edition) (World Scientific Lecture Notes in Physics).
    World Scientific Publishing Co. Pte. Ltd. 2005

\bibitem{parallelru} Vl.V. Voevodin, S.A. Zhumatiy,  S.I. Sobolev, A.S. Antonov, P.A. Bryzgalov, D.A. Nikitenko,
K.S. Stefanov, Vad.V. Voevodin  \emph{Practice of "Lomonosov"
Supercomputer}, \emph{Open Systems J.} - Moscow: Open Systems
Publ., 2012, no.7.

\end{thebibliography}
\end{document}